# Coordinating and Integrating Faceted Classification with Rich Semantic Modeling

Robert B. Allen[1][0000-0002-4059-2587] and Jaihyun Park[2][0000-0001-6217-7192]

[1] New York, NY, USA
[2] Syracuse University, Syracuse NY, USA
rba@boballen.info, jpark71@syr.edu

**Abstract.** Faceted classifications define dimensions for the types of entities included. In effect, the facets provide an "ontological commitment". We compare a faceted thesaurus, the Art and Architecture Thesaurus (AAT), with ontologies derived from the Basic Formal Ontology (BFO2), which is an upper (or formal) ontology widely used to describe entities in biomedicine. We consider how the AAT and BFO2-based ontologies could be coordinated and integrated into a Human Activity and Infrastructure Foundry (HAIF). To extend the AAT to enable this coordination and integration, we describe how a wider range of relationships among its terms could be introduced. Using these extensions, we explore richer modeling of topics from AAT that deal with Technology. Finally, we consider how ontology-based frames and semantic role frames can be integrated to make rich semantic statements about changes in the world.

## 1 Introduction

We have proposed the development of digital libraries with direct representation [4]. For instance, we have proposed developing "community models" to support the indexing of digitized historical newspapers. These models would represent complex material as highly-structured dynamic knowledgebases and semantic models.

Here, we explore developing the Human Activity and Infrastructure Foundry (HAIF) [8], which would be a collection of resources useful for describing common human activities and infrastructures such as those described in newspapers. Specifically, we explore developing the HAIF foundry by extending existing thesauri. We focus on the Art and Architecture Thesaurus (AAT) [17] [1] because its faceted structure facilitates recasting it as a structured, rich semantic

---

[1] http://www.getty.edu/research/tools/vocabularies/aat/

ontology. Potentially, this results in a resource with a wide range of applications.

HAIF would be comparable to the Open Biomedical Ontology (OBO) Foundry[2] [20] and could be interwoven with it. The OBO Foundry is a collection of biomedical ontologies which are generally consistent with the structure defined by the Basic Formal Ontology (BFO2) [10]. An upper (or formal) ontology such as BFO2 specifies the types of entities that can be used in reference ontologies as well as the relationships among them. Thus, for instance, BFO2 specifies that some entities should be Continuants (endurants) and others Occurrents (perdurants). In turn, Continuants are divided into Independent Continuants (e.g., Objects) and Dependent Continuants (e.g., Qualities, Disposition). BFO2 is a realist system based on Aristotelian principles; thus, in addition to proposing a hierarchy of Entity types, BFO2 also distinguishes between Universals and Particulars. In BFO2, an ontology is a single-inheritance taxonomy, a collection of "Is_A related" Entities.

## 2  Faceting, PMEST, and BFO2

Faceting describes different aspects or types of components of an Entity. Working with a faceted classification system may be analogized to viewing a gemstone through the perspectives of its facets. A few facet systems are universal; they cover all entities. Universal faceted systems are closely related to upper ontologies. As Svenonius [22] notes, facets provide "ontological commitment".

The best known universal system is Raganathan's PMEST, which is the basis of his Colon Classification [18]. Table 1 shows PMEST's similarity to some of BFO2's top-level Entity types. There are also differences and ambiguities. For instance, in BFO2 Parts are simply Independent_Continuants connected by Part_of relationships, whereas in PMEST Parts belong to a separate facet, the Matter or Property facet.

| PMEST | PMEST Description | BFO2 |
|---|---|---|
| **Personality** | Focal Subject | Independent Continuants |
| **Matter or Property** | Properties or Materials of the Subject. | Dependent Continuants and Parts |
| **Energy** | Processes | Processes |
| **Space** | Location | Spatial Regions, Sites |
| **Time** | Temporal Relationships | Temporal Regions |

**Table 1.** A rough comparison of the top level of PMEST and BFO2.

As noted above, many biomedical ontologies based on BFO2 have been collected into the OBO Foundry. Although efforts have been made to coordinate and standardize these OBO ontologies, some of the OBO ontologies do not strictly adhere to the BFO2 policies.[3] For instance, some Entities are subdivided by Parts, and those Parts are then further subdivided as "Is_A" taxonomies in a fashion similar to guide terms. Moreover, some OBO ontologies use combinations of terms in a manner similar to the pre-coordination of terms familiar from other controlled vocabularies. One recent effort for OBO proposes adopting cross-products [15] to improve coordination across multiple ontologies. Nonetheless, as with faceted classifications, compound terms and multiple inheritance remain issues in coordinating the OBO ontologies.

---

[2] http://obofoundry.org/

In contrast to SKOS[4], which has been described as implementing "weak semantics" [15] with simple linked-data approaches, we recognize several levels of rich semantics. As described in Section 5, we go beyond BFO2 to statements which describe changes in a dynamic Reality. Our approach uses rich semantics to describe Entities with structured schemas and supports descriptions of the interaction of those Entities.

## 3   Rich Semantic Faceting of AAT

The Getty Research Institute has developed a collection of resources for cultural informatics. The best known of these resources is AAT, which is a faceted thesaurus of generic terms associated with art and architecture that goes beyond the narrow sense of art and architecture to include diverse items such as Vehicles and Generators.[5] As a thesaurus, AAT focuses on Broader_Than (BT) and Narrower_Than (NT) relationships[6] as well as Related Terms (RT). The AAT hierarchy is carefully constructed and curated. The top two levels of the AAT hierarchies are shown in Table 2. In some cases (e.g., Materials), there are no subsections. In other cases, the hierarchies are deep; for instance, AAT:Object_Genres includes a wide range of items such as balls, silverware, and shutters. Recently, AAT has been formulated as linked data using SKOS. This includes refinement of the BT/NT relationships but is still short of the rich, structured relationships of the BFO2-based ontologies.

The AAT facets generally follow PMEST as extended by the Bliss Bibliographic Classification (BC2). The BC2 is a universal facet system which builds on PMEST by adding facets such as Agent, Patient, Operation, Product, and By-Product. While PMEST has Processes ("Energy"), BC2 extends those processes by considering the participants (i.e., Agent), the mechanism (i.e., Operation), and the things that changed (i.e., Product, By-Product). The extensions of PMEST by BC2 are compatible with the XFO programming environment we have developed as an extension of BFO2 [5]. Specifically, we interpret the extensions to BC2 as reflecting semantic roles.

---

[3] http://obofoundry.org/principles/fp-000-summary.html
[4] Simple Knowledge Organization System, https://www.w3.org/2004/02/skos/
[5] In the Aristotelian terminology of BFO2, AAT mostly describes Universals.
[6] In AAT, in some cases BT and NT are further specified into Part/Whole, Instance_Of, or Genus/Species relationships.

If PMEST were a full upper ontology, AAT could be formally descended from it. As an alternative, we consider integrating AAT's structure with BFO2/XFO. To merge AAT with BFO2, it will be necessary to restructure some of the categories. For instance, both AAT:Objects and AAT:Agents would be included as BFO2:Independent_Continuants, but should be distinguished because Agents operate on Objects (also see Section 5). For instance, a Person (Agent) Resides (Activity) in a House (Object).[7] In addition, there are nuanced distinctions between AAT and BFO2 that would need to be resolved. For instance, AAT:Built_Environments includes gardens and landscapes, but it is unclear whether a Landscape is well enough specified to be considered a BFO2:Object. Another example concerns AAT:Materials; in BFO2, there is no separate sub-entity for Substances (cf., [5, 23]).

| FACETS | SUB-CLASSES AND NOTES |
|---|---|
| Objects | Built Environments, Furnishings, and Equipment, Visual and Verbal Communications, Object Groupings and Systems, Object Genres, Components |
| Agents | People, Organizations, Living Organisms |
| Activities | Disciplines, Functions, Events, Physical and Mental Activities, Processes and Techniques |
| Physical Attributes | Attributes and Properties, Conditions and Effects, Design Elements, Color |
| Materials | Note: Flat list of materials |
| Styles/ Periods | Note: List of descriptors, many of which are associated with geographic areas |
| Associated Concepts | Note: A disjoint set of concepts that do not fit into the other facets. Not itself a true facet (see [Sorgel]) |

**Table 2.** Top two levels of AAT.[8]

## 4 Extending AAT Coverage of Technologies

As noted above, AAT has broad coverage beyond what is usually considered art and architecture. For instance, one large section is devoted to Technologies. There is an extensive list of Technological Objects in AAT such as Engines, Calculating Machines, and Light Bulbs; a range of Technology Agents (e.g., Inventors, Engineers, and Builders); and a range of Technological Activities (e.g., Printing, Silver Smithing). However, there are only weak structures in AAT to describe how such Objects, Agents, and Activities are interrelated. The relationship between Objects, Agents, and Activities could be framed as TICs coordinated with Semantic Roles (see Section 5).

---

[7] Potentially, some of this structure could be recovered from the Related Terms in AAT and from text analysis of the definitions for some of the terms. However, some useful links appear to be difficult to extract. For instance, AAT provides at best a very indirect association between "milking" and "dairy farms".

[8] A Brand Names facet was recently added. It is not relevant for this analysis and is not discussed.

Thus, Technologies may be explained by Mechanisms [14] which describe how the Technology works.[9] For example, the Mechanism of an electric light bulb would describe the flow of electrons through a titanium wire causing it to glow. We have discussed a structured specification for Mechanisms in [5].[10] Moreover, many Technologies are parts of Systems. A System is a group of interdependent Entities [4, 5] which interact via Mechanisms. We leave for future work the aspects that need to be defined and the deeper questions about how technology interacts with social structures. In the meantime, much can be accomplished in structuring descriptions of Technologies by focusing on the more straightforward initial questions.

## 5    Semantic Roles and Semantic Modeling with XFO

In [5], we proposed that ontologies should be extended by adding a Model-Layer. That is, they should include features such Thick Independent Continuants (TICs) and Transitionals. TICs are structured frames with required rather than ad hoc collections of elements [5]. A TIC is like a class in an object-oriented programming language; it provides a package of inter-related elements. In order to specify TICs, there needs to be a clearer specification of different types of Part_Of relationships, systematic explanations for multi-granular interactions of Entities, and a clearer definition of BFO2:Object_Aggregates.

We have noted the limited coverage of Occurrents in ontologies in general (e.g., [1]) and for BFO2 in particular (e.g., [3]). A modeling approach should also include Transitionals which specify the changes which they describe; we have used the verbs in FrameNet [2, 13][11]. FrameNet describes the semantic case roles associated with each of the verbs. The required Semantic Roles form a frame (or schema). Such Semantic Roles are related to, but distinct from, ontological categories such as those in BFO2. The combination of TICs and Semantic-Role schemas[12] provides a second layer in the Model Layer over the base ontology-level relationships.

---

[9] CIDOC-CRM is Event-based and potentially could be adapted to cover a broad range of Events. In the current versions, Events are mostly limited to content management.

[10] The distinction between processes models and class-diagrams is also found in UML, which is closely related to object-oriented programming. A Mechanism could be considered as an Information Artifact (e.g., a plan [12]), as an executable model (e.g., [5]), or as both.

[11] Other systems of verb frames (e.g., [19]) could be used.

[12] Potentially, these schemas could be implemented as a special type of named graph.

As we noted for AAT (see Section 3), and in both FrameNet and BC2, "agent" is a Semantic Role. An agent would be a BFO2:Independent_Continuant (or potentially a TIC). But, what distinguishes an Independent Continuant that is an implementer of a transition from an Independent_Continuant that is the recipient ("patient" in the context of medicine) of an action? Thus, it is clear that "agent" and "patient" both belong in the Model Layer.

Agency provides a type of causation (e.g., "The rock caused the window to break"). But, we believe that a complete account would define causation as linking one transition to another [3] (e.g., "The falling rock caused the window to break." In applications, we need to distinguish between claims of causation with respect to Particulars (i.e., histories) and claims about model-layer statements.

## 6   AAT as a Foundation for HAIF

Rather than developing HAIF from scratch, a more direct strategy would base it on AAT[13]. The AAT hierarchy is roughly analogous to a Foundry. The entries in many of the AAT taxonomies are subdivided by "guide terms", which effectively create sub-facets[14]. These sub-facets might be compared to the multiple thesauri in a Foundry. One example of the use of AAT guide terms is to organize Entities "by function". If the Entities were coded by BFO2:Function, they could be integrated directly into the ontology, rather than being treated as exceptions. As another example, the guide term "by specific context" is similar to the BFO2:Site entity type. The pairing of AAT:Objects with appropriate AAT:Activities would be a key benefit of extending AAT as a BFO2-style ontology or Foundry. Moreover, we can specify Mechanisms and Procedures associated with an Object. For instance, in [6] we proposed a semantic description of the workflow for making celadon pottery.

Potentially, intangible cultural heritage such as music or dance [8] could be covered as BFO2:Processes. A challenge for the application of BFO2 to cultural heritage is that cultural heritage includes non-historical mythological and religious (i.e., non-realist) characters and events. This may need to be reconciled with the realist foundation of BFO2. Perhaps the "real" entity for iconographic material is people's cognitions. Thus, the "objects" could be neural cell assemblies and mental models.[15]

---

[13] In addition to AAT, there are other thesauri and controlled vocabularies which could be helpful for developing HAIF. These include thesauri about agriculture and clothing and fashion. There are, also, some comprehensive cross-domain thesauri related to human activities and infrastructures such as the Social, History, and Industry Classification, EUROVOC, and the UNESCO thesaurus. However, because they are faceted to varying degrees, they would need additional development work to be coordinated with HAIF.

[14] See Zeng's discussion of microthesauri: https://www.getty.edu/research/tools/vocabularies/microthesauri_zeng.pdf

Beyond AAT, there are several other Getty Research Institute resources. For instance, there are resources which describe specific people (e.g., ULAN) and places (e.g., TGN); these resources also include limited hierarchies. Taken together the Getty vocabularies allow institutions to create metadata records about specific works of art such as with Categories for the Description of Works of Art (CDWA) and about other cultural objects with Cataloging Cultural Objects (CCO).[16] Because these vocabularies generally describe Particulars (i.e., instances of Universals), they could also be covered by a BFO2-based HAIF.

## 7  Toward an Extended Human Activity and Infrastructure Foundry (XHAIF) and Applications

HAIF will need to be developed in several steps.[17] The first would be to incorporate coverage from existing thesauri. A second would develop and extend the base terms as dynamic, semantic models, the Model Layer. A third step would use statements to develop and build knowledge about the world.

All three steps in developing HAIF could be used to implement and extend Community and Cultural Models. We originally conceived of Community Models as descriptions of specific communities described in historical newspapers [8]. Potentially, the entity-event fabric could be part of a broader "unified temporal map" of history [7]. In addition to structured descriptions for towns described in historical newspapers, a temporal map could incorporate the history of works of art and cultural objects. Evidence for claims could link back across the temporal map to earlier events. As coverage is increased, some of the specific Cultural and Community Models could be developed as "exemplars".[18]

AAT is an especially useful foundation for HAIF because it deals mostly with physical objects (e.g., buildings, furniture, etc.), particularly if its coverage of Technologies is extended. Broader Community and Cultural Models must also incorporate significantly abstract domains such as business, government, and law. While some aspects of these additional areas deal with physical objects, other aspects involve social ontology [9, 24].

---

[15] Iconclass is an organization system for mythological and religious objects. http://www.iconclass.org/

[16] http://www.getty.edu/research/tools/vocabularies/cco_cdwa_for_museums.pdf

[17] While in this paper we focus on the development of HAIF, in other papers we have applied rich semantics to developing highly-structured scientific research reports. Potentially, HAIF could also be applied to develop highly-structured social science research reports.

[18] Because there are so many species, biologists often focus their work on "model organisms", or "exemplars". Naturally, their ontologies also focus on those organisms, although there are challenges in coordinating ontologies across species [16]. By analogy, we might focus on developing "model communities". In distinction to exemplars, models of prototypical communities would be developed in the Model Layer. Such prototypes could then be specialized as models of specific communities. Of course, we need clear criteria for determining when a prototype could be inferred.

Although complex OWL-based inferences are often computationally impractical, the potential for inference should be explored for structured descriptions of communities and cultures. Potentially, inferences across a semantic model could draw connections between events and suggest bridges over gaps in historical knowledge. Moreover, as new claims are made, they could be validated against prior knowledge.

## 8 Conclusion

We believe it is possible and desirable to develop a unified approach to rich semantics. BFO2 may be viewed as a faceted classification system analogous to PMEST. By building on the faceted AAT that generally follows PMEST, a foundry may be developed for human activities and infrastructures which is consistent with BFO2. Moreover, the basic foundry ontologies may be extended to be a framework for semantic modeling by incorporating semantic roles. The resulting combination of ontological frames and semantic role (i.e., case) frames can be used for modeling complex scenarios.

## References

1. Allen, R.B.: Model-Oriented Information Organization: Part 1, The Entity-Event Fabric, D-Lib Magazine, July 2013, http://doi.org/10.1045/july2013-allen-pt1
2. Allen, R.B.: Frame-based Models of Communities and their History. *Histoinformatics*, LNCS 8359, Jan. 2014, 110-119, DOI:10.1007/978-3-642-55285-4_9
3. Allen, R.B., Chu, YM.: Architectures for Complex Semantic Models, IEEE Conference on Big Data and Smart Computing, Feb. 2015, 254-261, doi: 10.1109/35021BIGCOMP.2015.7072809
4. Allen, R.B., From Ontology to Structured Applied Epistemology, Rich Semantics and Direct Representation for Digital Collections, Tsukuba, Nov. 2016. ARXIV 1610.07241v2
5. Allen, R.B., Jones. T.K.: XFO: Toward Programming Rich Semantic Models, Apr. 2018, ARXIV:1805.11050
6. Allen, R.B., Kim, YH.: Semantic Modeling with Foundries, Jan. 2018, ARXIV 1801.00725
7. Allen, R.B., Song, H., Lee, B.E., Lee, J.Y.: Describing Scholarly Information Resources with a Unified Temporal Map, ICADL, Dec. 2106, 212-217, DOI: 10.1007/978-3-319-49304-6_25
8. Allen, R.B., Yang, E., Timakum, T.: A Foundry of Human Activities and Infrastructures, ICADL, Dec. 2017, 57-64, DOI:10.1007/978-3-319-70232-2_5
9. Almeida, M., Brochhausen, M., Silva, F.B., dos Santos, R.B.M.: Ontological Approach to the Normative Dimension of Organizations: an Application of Documents Acts Ontology, Ciência da Informação, 46(1), 214-227, Jan 2017, DOI: 10.18225/ci.inf..v46i1.4024

10. Arp, R., Smith, B., Spear, A.D.: Building Ontologies with Basic Formal Ontology, 2015, Cambridge MA, MIT Press
11. Baker, T., Sutton, S.: Linked Data and the Charm of Weak Semantics: Introduction: The Strengths of Weak Semantics, ASIST Bulletin, 41(4), 2015, 10-12.
12. Chu, YM., Allen, R.B.: Formal Representation of Socio-Legal Roles and Functions for the Description of History, TPDL, Sep. 2016, DOI: 10.1007/978-3-319-43997-6_30
13. Fillmore, C.J., Baker, C.: A Frames Approach to Semantic Analysis, In: The Oxford Handbook of Linguistic Analysis, B. Heine and H. Narrog (eds.), Oxford University Press, 2009, DOI: 10.1093/oxfordhb/9780199544004.013.0013
14. Machamer, P.K., Darden, L. Craver, C.F.: Thinking about Mechanisms, Philosophy of Science, 2000, 67, 1–25
15. Mungall, C.J., Bada, M., Bernardini, T.Z., Deegan, J., Ireland, A., Harris, M.A., Hill, D.P., Lomax, J.: Cross-Product Extensions of the Gene Ontology, Journal of Biomedical Informatics, 2011, 44(1):80-6, DOI: 10.1016/j.jbi.2010.02.002
16. Mungall, C.J., Torniai, C., Gkoutos, G.V., Lewis, S.E., Haendel, M.A.: Uberon: An Integrative Multi-Species Anatomy Ontology. Genome Biology. 2012, 13(1), R5, DOI: 10.1186/gb-2012-13-1-r5
17. Petersen, T., Barnett, P. (eds.): Guide to Indexing and Cataloging with the Art & Architecture Thesaurus. (2nd ed.) New York. Oxford University Press, 1994
18. Ranganathan, S.R.: Colon Classification (6th ed.) 2006 (1st ed. 1937), Ess-Ess Pub., Bangalore
19. Schuler, K.K.: VerbNet: A Broad-Coverage, Comprehensive Verb Lexicon, 2005, Ph.D. Dissertation, Department of Computer Science, University of Pennsylvania, repository.upenn.edu/dissertations/AAI3179808/
20. Smith, B., Ashburner, M., Rosse, C., Bard, J., Bug, W., Ceusters, W., Goldberg, L.J., Eilbeck, K., Ireland, A., C.J.: The OBI Consortium, Leontis, N., Rocca-Serra, P., Ruttenberg, A., Sansone, S-A., Scheuermann, R.H., Shah, N., Whetzel, P., Lewis,S.: The OBO Foundry: Coordinated Evolution of Ontologies to Support Biomedical Data Integration, Nature Biotechnology, 25 (11), 2007, 1251–1255, DOI:10.1038/nbt1346
21. Sorgel, D.: The Art and Architecture Thesaurus (AAT): A Critical Appraisal, Visual Resources 10(4), 1995, 369-400, DOI:10.1080/01973762.1995.9658306
22. Svenonius, E.: Definitional Approaches in the Design of Classification and Thesauri and Their Implications for Retrieval and Automatic Classification. In: Knowledge Organization for Information Retrieval (Ed.) L.C. McIlwaine, The Hague, International Federation for Information and Documentation, 1997, 12-16
23. Vogt , L., Grobe, P., Quast, B., Bartolomaeus, T.: Accommodating Ontologies to Biological Reality—Top-Level Categories of Cumulative-Constitutively Organized Material Entities, (January 9, 2012) PLOSOne, DOI: 10.1371/journal.pone.0030004
24. Zaibert, L., Smith, B.: The Varieties of Normativity: An Essay on Social Ontology, In S.L. Tsohatzidis (ed.), Intentional Acts and Institutional Facts: Essays on John Searle's Social Ontology. Springer, 57-173, 2007